\documentclass[twocolumn ,aps,showpacs,preprintnumbers,amsmath,amssymb]{revtex4}

\usepackage{amsthm}
\usepackage{graphics}
\usepackage[dvips]{graphicx}

\newcommand{\beq}{\begin{equation}}
\newcommand{\eeq}{\end{equation}}
\newcommand{\bcen}{\begin{center}}
\newcommand{\ecen}{\end{center}}
\newcommand{\bdm}{\begin{displaymath}}
\newcommand{\edm}{\end{displaymath}}
\newcommand{\best}{\begin{equation*}}
\newcommand{\eest}{\end{equation*}}
\newcommand{\bea}{\begin{eqnarray}}
\newcommand{\eea}{\end{eqnarray}}
\newcommand{\ot}{\otimes}

\newcommand{\locc}{\textrm{LOCC}}

\newcommand{\Ee}{E_\epsilon}

\newcommand{\ket}[1]{| #1 \rangle}

\newcommand{\dtr}{D_{\textrm{tr}}}

\renewcommand{\epsilon}{\varepsilon}
\renewcommand{\phi}{\varphi}

\newcommand{\tr}{{\textrm{Tr}}}
\newcommand{\Tr}{\tr}

\newcommand{\mcs}{{\mathcal{S}}}

\newtheorem{prop}{Proposition}

\newtheorem{cor}{Corollary}
\newtheorem{???}{Question}
\newtheorem{defi}{Definition}
\newtheorem{teo}{Theorem}
\newtheorem{lemma}{Lemma}
\newtheorem{dim*}{Proof}

\newtheorem{reason*}{Motivation for the question}

\begin{document}

\title{Epsilon-measures of entanglement}

\author{Caterina-E. Mora$^{1,2}$, Marco Piani$^{1,3}$, Hans-J. Briegel$^{2,3}$}
\affiliation{$^1$ Institute for Quantum Computing \& Department of Physics and Astronomy, University of Waterloo, University Ave. W., N2L 3G1, Canada\\
$^2$ Institut f\"ur Quantenoptik und Quanteninformation der \"Osterreichischen Akademie der Wissenschaften, Innsbruck, Austria\\
$^3$ Institut f{\"u}r Theoretische Physik, Universit{\"a}t Innsbruck, Technikerstra{\ss}e 25, A-6020 Innsbruck, Austria}

\begin{abstract}
We associate to every entanglement measure a family of measures which depend on a precision parameter, and which we call \emph{$\epsilon$-measures of entanglement}. Their definition aims at addressing a realistic scenario in which we need to estimate the amount of entanglement in a state that is only partially known. We show that many properties of the original measure are inherited by the family, in particular weak monotonicity under transformations applied by means of Local Operations and Classical Communication (LOCC). On the other hand, they may increase on average under stochastic LOCC.  Remarkably, the $\epsilon$-version of a convex entanglement measure is continuous even if the original entanglement measure is not, so that the $\epsilon$-version of an entanglement measure may be actually considered a smoothed version of it.
\end{abstract}

\maketitle

\section{Introduction}
\label{sec:intro}

Entanglement~\cite{reviewent} is a property of quantum states of two or more systems that 
allows for correlations that are stronger than those possible in classical physics. Entanglement has been shown to be a useful resource in the field of quantum information, in that it allows one to perform certain tasks in an enhanced way, i.e. more efficiently (as in quantum computing), 
more securely (as in quantum cryptography) or with a smaller amount of communication~\cite{NC}. 

Given the nature of entanglement as a resource, both its detection and its quantification are problems of fundamental relevance in quantum information. Deciding whether a state is entangled or not is, in general, difficult and over the years different methods have been devised to achieve this task~\cite{reviewent}. The quantification of entanglement is instead obtained by means of so-called entanglement measures. Different such measures exist in literature, 
and each is related to some particular aspect of entanglement~\cite{reviewent,reviewplenio}.

In the following we show how it is possible to associate to every entanglement measure a whole family of measures which we call $\epsilon$\emph{-measures} of entanglement and which depend on a precision parameter $\epsilon$. The motivation for studying $\epsilon$-measures of entanglement is three-fold.

First, their definition aims at addressing a realistic scenario in which we need to estimate the amount of entanglement in a state that is only partially known. On the one hand, this happens for the imperfect preparation of a target state $\rho$. Indeed,  any preparation apparatus has realistically only a certain degree of precision and reliability. On the other hand,  when we test what is the output state of the preparation procedure, even when we have a  good estimate $\rho$ of the state $\rho_{\textrm{true}}$ actually prepared (e.g. by  having done tomography on a finite amount of copies of the state), we are dealing with a certain degree of uncertainty on its parameters. One can then interpret the $\epsilon$-version $E_\epsilon$ of an entanglement measure $E$ as an estimate -- actually, a lower bound -- of the entanglement (as measured by $E$) that is for sure present in the system that we have prepared in a state $\rho$ with some approximation $\epsilon$. The mathematical definition of $\epsilon$-measure will correspondingly aim at quantifying the minimum ``guaranteed'' amount of entanglement, given the promise that the state which has actually been prepared is within a distance $\epsilon$ from some state $\rho$.

A second motivation  for studying this class of measures is that, as we will see, the $\epsilon$-version of a convex entanglement measure is continuous even if the original entanglement measure is not, so that the $\epsilon$-measures may be considered as a smooth version of the original ones~\footnote{In this sense, our approach is similar to the one adopted to define smooth min- and max-entropies as generalizations of von Neumann entropy in~\cite{smoothent,thesisrenner}.}.

Third, $\epsilon$-measures constitute a playground where to find entanglement measures that, e.g., satisfy some fundamental properties, as weak monotonicity under Local Operations and Classical Communication (LOCC), but not necessarily other ones, e.g. monotonicity on average (see below for definitions).

Moreover, on the one hand, one may hope that looking at the properties of the $\epsilon$-version of a certain entanglement measure, could lead to some insight about the latter; on the other hand, useful properties -- e.g., for assessing the universality of certain classes of states~\cite{univprl,univlong}-- may be inherited by the $\epsilon$-version of a certain entanglement measure, and while the latter may require a certain care because of, e.g., its discontinuity, the $\epsilon$-version may be handled more easily.

The paper is organized as follows. After giving the definition of $\epsilon$-measures in Section \ref{sec:def}, in Section \ref{sec:prop} we discuss their properties -- in particular, how they are related to the properties of the original measures. In Section \ref{sec:bounds} with provide bounds which establish a relation with distance-based entanglement measures. Section \ref{sec:conclusions} is devoted to our conclusions.

\section{Definition}
\label{sec:def}

\subsection{Entanglement monotones and entanglement measures}
\label{sec:monomeas}

Here and in the following, where not otherwise specified, we will consider finite-dimensional multipartite systems. We will denote the relevant set of states (density matrices)  by $\mathcal{D}$, and the subset of separable states by $\mcs$. The latter is given by those states which can be written as $\sum_k p_k \rho_k^{A_1}\ot \rho_k^{A_2}\ot\cdots\ot\rho_k^{A_2}$, with $p_k$ a probability distribution. Both $\mathcal{D}$ and $\mcs$ share the fundamental property of being convex.

We recall  now some properties that a function $E:\mathcal{D}\ni\rho\mapsto E(\rho)\in\mathbb{R^+}$, candidate to be an entanglement measure, may be asked to satisfy~\cite{reviewent,reviewplenio}:
\begin{description}
\item[{[VOS]}] $E$ vanishes on any separable state;
\item[{[LU]}] $E$ is invariant under local unitaries;
\item[{[WEM]}] weak monotonicity, i.e. $E$ is non-increasing under LOCC operations: $E(\Lambda_{\textrm{LOCC}}[\rho])\leq E(\rho)$;
\item[{[MOA]}] monotonicity on average, i.e. $E$ does not increase \emph{on average} under LOCC operations: $E(\rho)\geq\sum_i p_i E(\rho_i)$, where $\rho_i$ are the possible outputs, each with probability $p_i$, of an LOCC operation;
\item[{[CO]}] $E$ is convex: $E(p\rho_1+(1-p)\rho_2)\leq pE(\rho_1)+(1-p)E(\rho_2)$.
\end{description}
Property [LU] is contained in [WEM], but typically listed separately in literature. Property [WEM] is weaker than [MOA], as it refers to the case where there is just one possible output. If a function $E$ satisfies property [WEM] we say that it is a \emph{weak entanglement monotone}. If it satisfies [MOA], we consider it as \emph{strong entanglement monotone}. Fulfilling also [VOS] promotes weak (strong) entanglement monotones to \emph{weak (strong) entanglement measures}. Convexity is an additional requirement, quite convenient from a mathematical point of view. It is often considered a necessary condition for a good entanglement measure, being somehow related to (classical) information loss~\cite{vidal}, though such a point of view has been questioned~\cite{plenio,reviewent}.

We will address one further property, that applies only to the case of candidate multipartite measures defined in the case where the number of parties is not fixed:
\begin{description}
\item[{[TE]}] trivial extensitivity: $E(\rho_{A_1\ldots A_N}\otimes\sigma_{A_{N+1}})=E(\rho_{A_1\ldots A_N})$, for any $N+1$-party state $\rho_{A_1\ldots A_N}\otimes\sigma_{A_{N+1}}$, i.e. such that one party is uncorrelated with the rest.
\end{description}
Property [TE] could be taken as part of [WEM], if we consider discarding or adding an uncorrelated party a local operation. We remark that [TE] may not be satisfied by entanglement measures tailored for a fixed number of parties. E.g., multipartite measures defined as averages of bipartite measures may not satisfy [TE] if the averaging is not suitably chosen~\cite{univlong}.

\begin{defi}[$\epsilon$-measure]
For every entanglement measure $E$, and any $\epsilon\geq0$ we define the associated $\epsilon$-measure (with respect to a distance $D$)
\beq
\label{eq:defeps}
E^{(D)}_\epsilon(\rho)=\min\{E(\sigma)~|~\sigma\in B^{(D)}_\epsilon(\rho)\},
\eeq
where the $B^{(D)}_\epsilon(\rho)\subseteq \mathcal{D}$ is the set of states $\sigma\in\mathcal{D}$ such that $D(\rho,\sigma)\leq\epsilon$, for some fixed distance measure $D$. 
\end{defi}
\noindent
We are allowed to consider a minimum, rather than an infimum, in Definition \ref{eq:defeps} because we are considering finite-dimensional systems for which the set of (separable) states is compact \cite{geometrystates}.

For the sake of simplifying notation, here and in the following we will omit the superscript $(D)$ where it does not give rise to misunderstandings. Obviously, $E^{(D)}_{\epsilon=0}(\rho)=E(\rho)$ for any distance $D$~\footnote{A distance $D$ satisfies: $D(\rho,\sigma)=0$ if and and only if $\rho=\sigma$.}, and $E^{(D)}_{\epsilon}(\rho)$ vanishes on all the set of states $\mathcal{D}$, if $\epsilon$ is so large that there is a separable state in $B^{(D)}_\epsilon(\rho)$. Thus, we have in mind the case in which $\epsilon$ is small but non-zero. In particular, for a fixed entangled state $\rho$, we are typically interested in values $\epsilon$ that are smaller than the distance of $\rho$ from the set of separable states. Moreover, we observe that while we refer to entanglement measures, the construction of Definition \ref{eq:defeps} could be applied to any functional $\mathcal{D}\rightarrow \mathbb{R}$ to obtain its $\epsilon$-version. 

The quantity $E^{(D)}_\epsilon$ clearly solves the task of quantifying the ``guaranteed'' entanglement, since, by definition, any state $\sigma$ within an $\epsilon$-distance of the desired state $\rho$ has $E(\sigma)\geq\Ee(\rho)$. 

\section{Properties}
\label{sec:prop}

In this section we prove some of the main properties of the $\epsilon$-measures of entanglement. In the following, we restrict ourselves to choices of distance measures that are convex in each entry separately~\footnote{A distance is symmetric: $D(\rho,\sigma)=D(\sigma,\rho)$, thus convexity in the first entry implies convexity in the second entry}:
\[
D(p\rho_1+(1-p)\rho_2,\sigma)\leq pD(\rho_1,\sigma)+(1-p)D(\rho_2,\sigma),
\]
or jointly convex:
\[
D(p\rho_1+(1-p)\rho_2,p\sigma_1+(1-p)\sigma_2)\leq pD(\rho_1,\sigma_1)+(1-p)D(\rho_2,\sigma_2).
\]
Furthermore, we will ask that such measures be contractive under completely positive and trace preserving (CPT) operations:
\[
D(\Lambda[\rho],\Lambda[\sigma])\leq D(\rho,\sigma).
\]
The latter requirement is natural and necessary in order to consider any distance measure as a measure of distinguishability~\cite{distinguish}. Moreover, it also allows us to make the following observation (valid also for smooth entropies \cite{thesisrenner}):
\begin{prop}
The minimum in \eqref{eq:defeps} can be restricted to states $\sigma$ whose support is contained in the tensor product of the local supports of $\rho$.
\end{prop}
\begin{proof}
Consider the following LOCC protocol. Each party $A_i$ performs a non-complete von Neumann measurement, given by the projection $P_{A_i}$ onto the local support of $\rho$, i.e. on the support of $\rho_{A_i}=\Tr_{\backslash A_i}(\rho)$, where $\Tr_{\backslash A_i}$ denotes the trace over all subsystems but $A_i$, and its complement $\openone_{A_i}-P_{A_i}$. If all the parties have obtained the result corresponding to $P_{A_i}$, they keep the resulting state
\[
\sigma_P=\frac{1}{p_\sigma}\Big(\bigotimes_{i=1}^n P_{A_i}\Big) \sigma \Big(\bigotimes_{i=1}^n P_{A_i}\Big),\qquad p_\sigma=\Tr\Big(\bigotimes_{i=1}^n P_{A_i} \sigma\Big),
\]
which by definition is contained in the tensor product of the local supports. Otherwise, they create, e.g., one separable state $\sigma_{\rm sep}$ by scratch.  This protocol is described by an LOCC map $\Lambda_{\locc}$ whose action on any state $\sigma$ is
\beq
\Lambda_{\locc}[\sigma]=p_{\sigma} \sigma_P + (1-p_\sigma) \sigma_{\rm sep}.
\eeq
Since the support of $\rho$ is contained in the tensor product of the local supports, $\Lambda_{\locc}[\rho]=\rho$. Thus, for any $\sigma\in B^{(D)}_\epsilon(\rho)$,
\[
D(\rho,\sigma)\geq D(\Lambda_{\locc}[\rho],\Lambda_{\locc}[\sigma])=D(\rho,\Lambda_{\locc}[\sigma]),
 \]
i.e. $\Lambda_{\locc}[\sigma]$ is also contained in $B^{(D)}_\epsilon(\rho)$. Moreover, as [WEM] holds for $E$, $E(\Lambda_{\locc}[\sigma])\leq E(\sigma)$.
\end{proof}

An example of jointly convex, contractive distance is given by the trace distance $\dtr(\rho,\sigma)=\frac{1}{2}\Tr |\rho-\sigma|$, with $|X|=\sqrt{X^\dagger X}$.

\subsection{Monotonicity}

First of all, we show that the $\epsilon$-generalization of any weak entanglement measure is again a weak measure.
\begin{teo}
Given any weak entanglement measure $E$, $\Ee$ is a weak entanglement measure.
\end{teo}
\begin{proof}
In order to prove the statement, we must prove that if $E$ satisfies the properties of Section \ref{sec:monomeas}, then they also hold for $\Ee$. 
\begin{description}
\item[{[VOS]}] In order to prove that $\Ee$ vanishes on separable states it is sufficient to consider that, for any $\rho_{\textrm{Sep}}$, we have $0\leq\Ee(\rho_{\textrm{Sep}})\leq E(\rho_{\textrm{Sep}})=0$.
\item[{[WEM]}]
Let us consider a state $\sigma\in B_\epsilon(\rho)$  which realizes the minimum in the definition of $\Ee$:
\beq
\Ee(\rho)=E(\sigma).
\eeq
Since $E$ is a weak monotone, we have $E(\sigma)\geq E(\Lambda_\locc[\sigma])$. Moreover, we have chosen the distance $D$ to be contractive under CPT maps, so that $\epsilon\geq D(\sigma,\rho)\geq D(\Lambda_\locc[\sigma],\Lambda_\locc[\rho])$. Therefore, $\Lambda_\locc[\sigma]\in B_\epsilon(\Lambda_\locc[\rho])$. It follows that 
\beq
\begin{split}
\Ee(\rho)&=E(\sigma)\geq E(\Lambda_\locc[\sigma])\\
	& \geq \min\{ E(\tau)~|~\tau\in B_\epsilon(\Lambda_\locc[\rho])\}\\
	&=\Ee(\Lambda_\locc[\rho]).
\end{split}
\eeq
\item[{[LU]}] Invariance under LU follows immediately from [WEM].
\end{description}
\end{proof}

As regards convexity, it is inherited if the distance used in the definition of $\Ee$ is jointly convex.
\begin{teo}
Given any convex entanglement monotone $E$, $\Ee=E^{(D)}_\epsilon$ is convex if the distance $D$ is jointly convex.
\end{teo}
\begin{proof}
Considering states $\sigma_1\in B_\epsilon(\rho_1),\,\sigma_2\in B_\epsilon(\rho_2)$ such that 
\beq
\Ee(\rho_1)=E(\sigma_1),\qquad \Ee(\rho_2)=E(\sigma_2),
\eeq
we have
\beq
\begin{split}
pE_\epsilon(\rho_1)+(1-p)\Ee(\rho_2)
&=
pE(\sigma_1)+(1-p)E(\sigma_2)\\
&\geq
E(p\sigma_1+(1-p)\sigma_2)\\
&\geq
\Ee(p\rho_1+(1-p)\rho_2),
\end{split}
\eeq
because $E$ is convex. Thanks to the joint convexity of $D$, we have then:
\beq
\begin{split}
D(p\rho_1&+(1-p)\rho_2, p\sigma_1+(1-p)\sigma_2)\\
	&\leq pD(\rho_1,\sigma_1)+ (1-p)D(\rho_2,\sigma_2)\\
	&\leq\epsilon.
\end{split}
\eeq
\end{proof}

We treat separately the [TE] property because the change in the number of parties is a very particular type of operation, and its inclusion in the set of LOCC operation could be debatable.
\begin{teo}
If a weak entanglement monotone $E$ satisfies [TE], then also $\Ee$ does.
\end{teo}
\begin{proof}
Since the distance $D$ is contractive under CPT maps, and in particular under partial trace, we have that, for any $\sigma_{A_1\ldots A_N A_{N+1}}, \tau_{A_1\ldots A_N A_{N+1}}$, $D(\sigma_{A_1\ldots A_NA_{N+1}},\tau_{A_1\ldots A_NA_{N+1}})\geq D(\sigma_{A_1\ldots A_N},\tau_{A_1\ldots A_N})$, where $\sigma_{A_1\ldots A_N}=\tr_{A_{N+1}}[\sigma_{A_1\ldots A_NA_{N+1}}]$
and $\tau_{A_1\ldots A_N}=\tr_{A_{N+1}}[\tau_{A_1\ldots A_NA_{N+1}}]$.
 As a consequence  of this, a reduced state $\sigma_{A_1\ldots A_N}$ is in $B_\epsilon(\tau_{A_1\ldots A_N})$ for all $\sigma_{A_1\ldots A_N A_{N+1}}\in B_\epsilon(\tau_{A_1\ldots A_{N}A_{N+1}})$. Therefore,
 \begin{widetext}
\beq
\begin{split}
\Ee(\rho_{A_1\ldots A_N}\otimes \tau_{A_{N+1}}) &	= \min\{ E(\sigma_{A_1\ldots A_N A_{N+1}})~|~\sigma_{A_1\ldots A_N A_{N+1}}\in B_\epsilon(\rho_{A_1\ldots A_N}\otimes \tau_{A_{N+1}})\}\\
	&\geq \min\{ E(\sigma_{A_1\ldots A_N A_{N+1}})~|~\sigma_{A_1\ldots A_N}\in B_\epsilon(\rho_{A_1\ldots A_N})\}\\
	&\geq \min\{E(\sigma_{A_1\ldots A_N}\otimes \tau_{A_{N+1}})~|~\sigma_{A_1\ldots A_N}\in B_\epsilon(\rho_{A_1\ldots A_N})\}\\
	&= \min\{E(\sigma_{A_1\ldots A_N}) ~|~\sigma_{A_1\ldots A_N}\in B_\epsilon(\rho_{A_1\ldots A_N})\}\\
	&= \Ee(\rho_{A_1\ldots A_N}).
\end{split}
\eeq
\end{widetext}
The first inequality comes from having enlarged the set of states over which the minimum is taken. The second one is due to the fact that, for any $\sigma_{A_1\ldots A_N A_{N+1}}$ and any fixed $\tau_{A_{N+1}}$, one has $E(\sigma_{A_1\ldots A_NA_{N+1}})\geq E(\sigma_{A_1\ldots A_N}\otimes \tau_{A_{N+1}})$, since there exists an LOCC operation (actually a local one) $\Lambda_{\textrm{LO}}: \sigma_{A_1\ldots A_NA_{N+1}}\rightarrow\sigma_{A_1\ldots A_N}\otimes \tau_{A_{N+1}}$.

On the other hand,
\begin{widetext}
\beq
\begin{split}
\Ee(\rho_{A_1\ldots A_N}\otimes \tau_{A_{N+1}})  & = \min\{E(\sigma_{A_1\ldots A_N A_{N+1}})~|~\sigma_{A_1\ldots A_N A_{N+1}}\in B_\epsilon(\rho_{A_1\ldots A_N}\otimes \tau_{A_{N+1}})\}\\
	&\leq \min\{E(\sigma_{A_1\ldots A_N}\otimes \tau_{A_{N+1}})~|~\sigma_{A_1\ldots A_N}\otimes \tau_{A_{N+1}}\in B_\epsilon(\rho_{A_1\ldots A_N}\otimes \tau_{A_{N+1}})\}\\
&= \min\{E(\sigma_{A_1\ldots A_N})~|~\sigma_{A_1\ldots A_N}\otimes \tau_{A_{N+1}}\in B_\epsilon(\rho_{A_1\ldots A_N}\otimes \tau_{A_{N+1}})\}\\
	&= \min\{E(\sigma_{A_1\ldots A_N})~|~\sigma_{A_1\ldots A_N}\in B_\epsilon(\rho_{A_1\ldots A_N})\}= \Ee(\rho_{A_1\ldots A_N}).
\end{split}
\eeq
\end{widetext}
The inequality is due to  the restriction of the set of states over which the maximum is taken. In the last equality we have used the fact that, for any state $X$, the mapping $X_{A_1\ldots A_N}\rightarrow X_{A_1\ldots A_N}\otimes \tau_{A_{N+1}}$ is a CPT map that can be reversed by tracing out system $A_{N+1}$, i.e. by another CPT map. Therefore, because of contractivity, $D(\rho_{A_1\ldots A_N}\otimes \tau_{A_{N+1}},\sigma_{A_1\ldots A_N}\otimes \tau_{A_{N+1}})=D(\rho_{A_1\ldots A_N},\sigma_{A_1\ldots A_N})$.
\end{proof}

From what we have seen, given any weak entanglement measure $E$, its $\epsilon$-generalization $\Ee$ is still a weak measure. Indeed,  by means of LOCC operations alone one cannot increase at all entanglement, as measured by the starting measure $E$. Moreover, the adopted distance is contractive, so that we can use neither LOCC nor non-LOCC operations to distinguish states better. As a consequence, we can not increase deterministically the guaranteed entanglement. Nevertheless, as we show below, it is possible to increase the guaranteed entanglement  of a state on average, even when $E$ is monotone on average. Physically, this might be interpreted as the fact that, when working in a probabilistic scenario, we may be able to guarantee the presence of entanglement under the condition of obtaining a certain result, even if we can not guarantee that it is always present.

\begin{prop}
$\Ee$ does not satisfy [MOA] -- monotonicity on average.
\end{prop}
\begin{proof}
Let us consider a state $\rho$ of the form: $\rho=\eta|0\rangle\langle0|\otimes\rho_{\rm Ent} + (1-\eta)|1\rangle\langle1|\otimes\rho_{\rm Sep}$ with  $\eta\in[0,1]$, where $\ket{0}$ and $\ket{1}$ are orthogonal states of a local ancilla, and where we have chosen $\rho_{\rm Ent}$ and $\rho_{\rm Sep}$ such that $\Ee(\rho_{\rm Ent})>0$ and $\Ee(\rho_{\rm Sep})=0$. For some choices of $\eta$, $\Ee(\rho)=0$: it is in fact sufficient to consider that the state $\sigma=|1\rangle\langle1|\otimes\rho_{\rm Sep}$ is separable and, for $\eta$ ``small enough'' (depending on the choice of distance $D$) $D(\sigma,\rho)\leq\epsilon$. If, for example, we consider the trace distance, then $\dtr(\rho,\sigma)=\eta \dtr(\rho_{\rm Ent}, \rho_{\rm Sep})\leq \epsilon$ for all $\eta\leq\epsilon/\dtr(\rho_{\rm Ent},\rho_{\rm Sep})$.

Consider now the local measurement that acts projecting on the single-qubit states $|0\rangle$ and $|1\rangle$ of the local ancilla. The possible outputs of such a measurement on $\rho$ are $\rho_{\rm Ent}$ (with probability $\eta$) and $\rho_{\rm Sep}$ (with probability $1-\eta$). We have thus:
\beq
\sum_i {p_i \Ee(\rho_i)} = \eta\Ee(\rho_{\rm Ent}) > 0 = \Ee(\rho).
\eeq
\end{proof}

The violation of [MOA] for $\Ee$ can be bounded in terms of the violation -- if present -- of [MOA] for $E$ and of the difference between $E$ and $\Ee$:
\beq
\begin{split}
\sum_i &{p_i  \Ee(\rho_i)} - \Ee(\rho) \leq \sum_i {p_i E(\rho_i)} - \Ee(\rho) \\
			&= \big[\sum_i {p_i E(\rho_i)} - E(\rho)\big] + \big[E(\rho)- \Ee(\rho)\big].
\end{split}
\eeq

We have shown that, with the exception of [MOA], the $\epsilon$-generalization of any measure is still a good entanglement measure. In the following we prove that other possibly relevant properties of the original quantity $E$, hold for $\Ee$ too. As a consequence, we expect that methods commonly used in literature to estimate, or bound, the entanglement of some states can be applied to the $\epsilon$-measure as well (see also Section \ref{sec:bounds})

\subsection{Continuity}

The $\epsilon$-generalization $\Ee^{(D)}$ of entanglement monotones has an advantage, in that it allows one to transform non-continuous measures (such as, for example, the Schmidt measure~\cite{schmidt} or the $\chi$-width~\cite{chiwidth}) into continuous ones. This results holds for any choice of distance $D$ and for all bounded and convex entanglement measures. The requirement that a measure be bounded is satisfied for finite-dimensional systems by all measures known to the authors, even when the measure is not continuous, e.g., for the Schmidt measure.

The following theorem is quite general, and applies not only to entanglement monotones/measures.
\begin{teo}
\label{thm:continuity}
Let $D$ be a convex distance measure on the set of states $\mathcal{D}$, and $E:\mathcal{D}\rightarrow\mathbb{R}^+$ be a convex and bounded functional. Then, $\Ee^{(D)}(\rho)$ is continuous in $\epsilon$ and $\rho$, for all $\epsilon>0$ and for all $\rho$.
\end{teo}
\begin{proof}
In order to prove the statement we shall prove separately the continuity in $\epsilon$ and $\rho$. 

(\emph{Continuity in $\epsilon$.}) We want to prove that, for any state $\rho$ and for any $\epsilon>0$, $|E_{\epsilon}(\rho)-E_{\epsilon'}(\rho)|\rightarrow 0$ as $\epsilon'\rightarrow \epsilon$. Let us fix $\rho$, and let $\epsilon_2=\max\{\epsilon,\epsilon'\}>0$ and $\epsilon_1=\min\{\epsilon,\epsilon'\}$.
Consider now  the state $\sigma_2$, $D(\sigma_2,\rho)\leq\epsilon_2$ such that $E_{\epsilon_2}(\rho)=E(\sigma_2)$. Moreover let us define $\lambda=\epsilon_1/\epsilon_2$ and $\sigma_\lambda=\lambda\sigma_2+(1-\lambda)\rho$, so that $D(\sigma_\lambda,\rho)\leq \frac{\epsilon_1}{\epsilon_2}D(\sigma_2,\rho)\leq \epsilon_1$ because of the convexity of the distance (see Fig.~\ref{fig:cont-eps}).
\begin{figure}[h]
\label{fig:cont-eps}
 \includegraphics[width=0.45\textwidth]{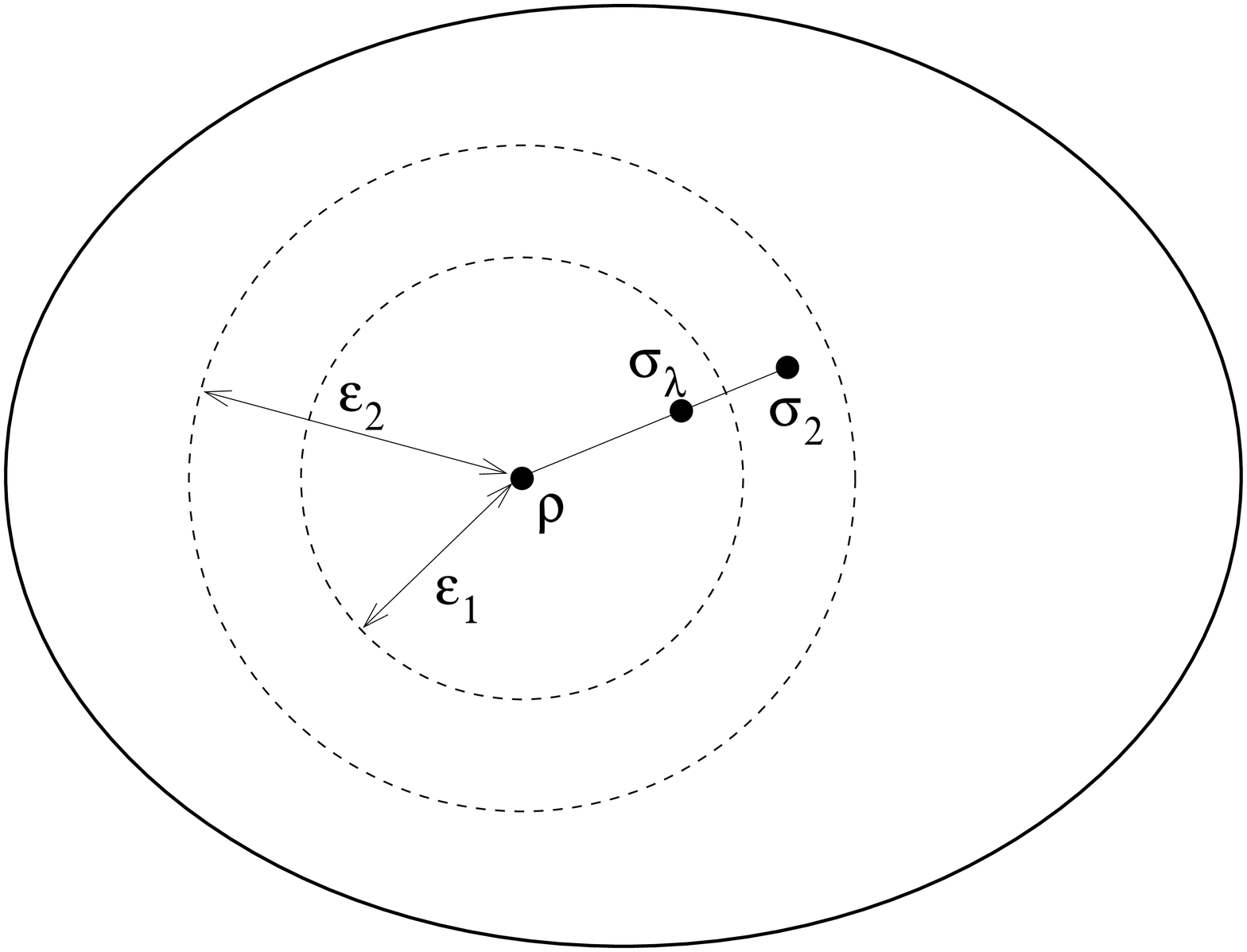}
\caption{Proof of the continuity in $\epsilon$ (for $\epsilon>0$) of $E_\epsilon(\rho)$. The ellipse denotes the whole set of states, while the two dashed circles represent the two balls of radius $\epsilon_{1,2}$ around $\rho$. See the main text for more details.}
\end{figure}
We have then,
\best
\begin{split}
E_{\epsilon_1}(\rho)&\leq E(\sigma_\lambda)\\
					&\leq \lambda E(\sigma_2)+(1-\lambda)E(\rho)\\
					&= \lambda E_{\epsilon_2}(\rho)+(1-\lambda)E(\rho),
\end{split}
\eest
which implies
\[
0\leq E_{\epsilon_1}(\rho)-E_{\epsilon_2}(\rho)\leq \frac{\epsilon_2-\epsilon_1}{\epsilon_2}(E(\rho)-E_{\epsilon_2}(\rho)).
\]

(\emph{Continuity in $\rho$.}) We want to prove that, for any $\epsilon>0$, and for any state $\rho_1$, $|\Ee(\rho_2)-\Ee(\rho_1)|\rightarrow0$ as $D(\rho_1,\rho_2)\rightarrow 0$. Let us fix $\epsilon>0$, and consider a state $\sigma_1\in B_\epsilon(\rho_1)$ such that $\Ee(\rho_1)=E(\sigma_1)$. We define now $\eta=D(\rho_1,\rho_2)$, and take $\lambda=\epsilon/(\epsilon+\eta)$  in $\sigma_\lambda=\lambda\sigma_1+(1-\lambda)\rho_2$ (see Fig.~\ref{fig:cont-rho}).
\begin{figure}[h]
\label{fig:cont-rho}
 \includegraphics[width=0.45\textwidth]{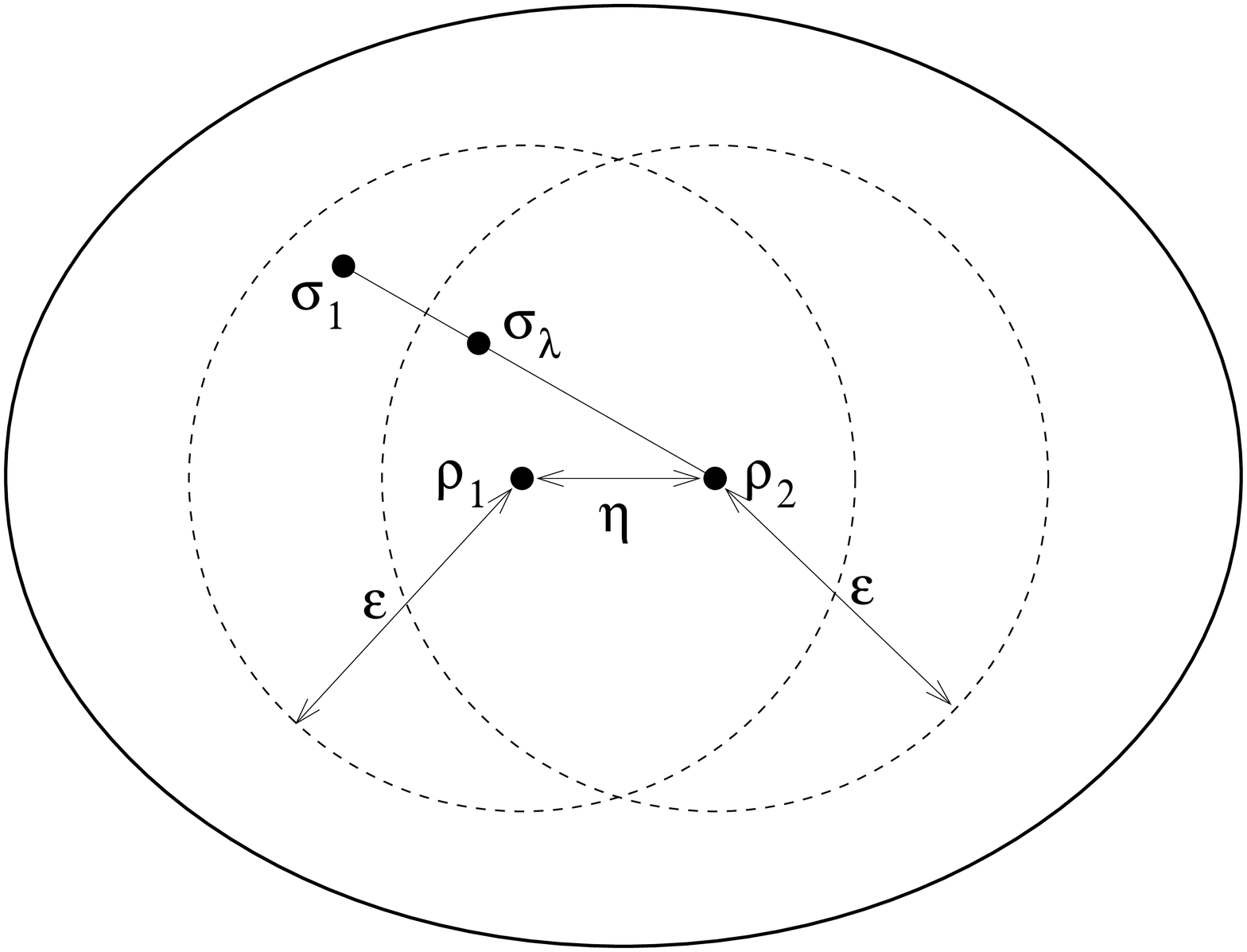}
\caption{Proof of the continuity in $\rho$ (for $\epsilon>0$) of $E_\epsilon(\rho)$. The ellipse denotes the whole set of states, while the two dashed circles represent the two balls of radius $\epsilon$ around $\rho_{1,2}$. See the main text for more details.}
\end{figure}
Then
\best
\begin{split}
D(\sigma_\lambda,\rho_2)&\leq \lambda D(\sigma_1,\rho_2)\\
	&\leq\lambda(D(\sigma_1,\rho_1)+D(\rho_1,\rho_2))\\
	&\leq\lambda(\epsilon+\eta)=\epsilon,
\end{split}
\eest
having used the  convexity of $D$ and the triangle inequality, in the first and in the second inequality, respectively. Therefore, $\sigma_\lambda$ is in $B_\epsilon(\rho_2)$ and
\best
\begin{split}
E_{\epsilon}(\rho_2)&\leq E(\sigma_\lambda)\\
					&\leq \lambda E(\sigma_1)+(1-\lambda)E(\rho_2)\\
					&= \lambda E_{\epsilon}(\rho_1)+(1-\lambda)E(\rho_2),
\end{split}
\eest
by the convexity of $E$ and the choice of $\sigma_1$.
Substituting the expression of $\lambda$ in terms of $\epsilon$ and $\eta$ and re-arranging the terms, we get
\[
E_{\epsilon}(\rho_2)-E_{\epsilon}(\rho_1)\leq \frac{\eta}{\epsilon+\eta}(E(\rho_2)-E_\epsilon(\rho_1)).
\]
Notice that $E(\rho_2)\geq E(\rho_2)-E_\epsilon(\rho_1)\geq 0$ for $\eta\leq\epsilon$.
Exchanging the role of $\rho_1$ and $\rho_2$, we get similarly
\[
E_{\epsilon}(\rho_2)-E_{\epsilon}(\rho_1)\geq - \frac{\eta}{\epsilon+\eta}(E(\rho_1)-E_\epsilon(\rho_2)),
\]
and again $E(\rho_1)\geq E(\rho_1)-E_\epsilon(\rho_2)\geq 0$ for $\eta\leq\epsilon$.
Taking $M\geq \max\{E(\rho_2)-E_\epsilon(\rho_1),E(\rho_1)-E_\epsilon(\rho_2)\}$, we then have
\[
|E_{\epsilon}(\rho_2)-E_{\epsilon}(\rho_1)|\leq\frac{\eta}{\epsilon+\eta}M.
\]
\end{proof}

\begin{cor}
Let $D$ be a convex distance measure, and $E:\mathcal{D}\rightarrow\mathbb{R}^+$ be convex and bounded. Then, if $E$ is continuous, $\Ee^{(D)}(\rho)$ is continuous in $\epsilon$ and $\rho$.
\end{cor}
\begin{proof}
In Theorem \ref{thm:continuity} we have shown that $\Ee(\rho)$ is continuous in $\epsilon$ and $\rho$ for all $\epsilon>0$. Continuity in $\rho$ for $\epsilon=0$ follows trivially from the fact that $E_0(\rho)=E(\rho)$, which is assumed to be continuous. What remains to be proved is the fact that $\Ee(\rho)$ is continuous in $\epsilon=0$ as a function of $\epsilon$, for all fixed $\rho$. More precisely, we want to prove that, for any state $\rho$ and for any $\eta>0$, there exists a $\mu>0$  such that
\beq
\label{epscont}
0\leq\epsilon\leq\mu\Rightarrow |E(\rho)-\Ee(\rho)|\leq\eta.
\eeq
Since $E(\rho)$ is continuous, we know that there exists a $\delta$ such that
\beq
\label{eqcont}
D(\sigma,\rho)\leq\delta\Rightarrow|E(\rho)-E(\sigma)|\leq\eta.
\eeq
This in particular must hold true for any $\sigma$ such that $E(\sigma)=\Ee(\rho)$, for all $\epsilon\leq\delta$. It follows thus that Eq. \eqref{epscont} is satisfied by choosing $\mu\leq\delta$. 
\end{proof}

We have thus shown that, for any $E:\mathcal{D}\rightarrow\mathbb{R}^+$, $\Ee(\rho)$ is always a continuous function of $\epsilon$ and $\rho$ for any $\epsilon>0$, and is continuous in $\epsilon=0$ whenever $E$ is continuous.

\subsection{Monogamy}


One important property of entanglement is that the same party, let us say $C$,  cannot be maximally entangled separately with different parties, let us say $A$ and $B$, at the same time, i.e. the two reduced states $\rho_{AC}$ and $\rho_{BC}$ cannot be both maximally entangled. On the other hand, non-maximal quantum correlations can exist between one party and different other parties, but the strength of such correlations will in general obey a trade-off relation. This was first proved in \cite{monogamy} using the concurrence \cite{concurrence} measure. This property is called \emph{monogamy} of entanglement, and it is formalized by means of inequalities of the form:
\beq
E^{A|C}(\rho)+E^{B|C}(\rho)\leq E^{(AB)|C}(\rho),
\label{monogamyineq}
\eeq
where $E$ is a \emph{bipartite} entanglement measure, $E^{A|C}$ ($E^{B|C}$) is the entanglement between systems $A$ and $C$ ($B$ and $C$), and $E^{(AB)|C}$ is the entanglement between the composite system $(AB)$ and $C$. Not all entanglement measures satisfy a monogamy inequality. We now show that if a measure $E$ admits a monogamy inequality, then $\Ee$ does too. 

\begin{teo}
If $E$ is an entanglement measure such that a monogamy inequality of the form \eqref{monogamyineq} holds, then the same inequality holds for $\Ee$.
\end{teo}
\begin{proof}
The proof follows straightforwardly from the definition of $\Ee$, in fact
\begin{widetext}
\beq
\begin{split}
\Ee^{(AB)|C}(\rho_{ABC}) & =\min\{E^{(AB)|C}(\sigma_{ABC})~|~\sigma_{ABC}\in B_\epsilon(\rho_{ABC})\}\\
	& \geq \min\big\{\big(E^{A|C}(\sigma_{AC})+E^{B|C}(\sigma_{BC})\big)~|~\sigma_{ABC}\in B_\epsilon(\rho_{ABC})\big\}\\
		& \geq \min\{E^{A|C}(\sigma_{AC})~|~\sigma_{ABC}\in B_\epsilon(\rho_{ABC})\} + \min\{E^{B|C}(\sigma_{BC})~|~\sigma_{ABC}\in B_\epsilon(\rho_{ABC})\}\\
	& \geq \min\{E^{A|C}(\sigma_{AC}) ~|~\sigma_{AC}\in B_\epsilon(\rho_{AC})\}  + \min\{ E^{B|C}(\sigma_{BC})~|~\sigma_{BC}\in B_\epsilon(\rho_{BC})\}\\
	&=  \Ee^{A|C}(\rho) + \Ee^{A|B}(\rho),
\end{split}
\eeq
\end{widetext}
where the three inequalities are respectively justified by: the monogamy of $E$; considering the sum of the minima rather than the minimum of the sum; enlarging the sets over which we take the minima.
\end{proof}

\section{Bounds and relation with distance measures}
\label{sec:bounds}

Let us consider the family of distance-based entanglement measures, as introduced in~\cite{VPRK1997, PlenioVedral1998}. If $\mcs$ is the set of separable states, and $D$ is a distance on the set of states one can define the quantity
\beq
\label{eq:distmeas}
E_{D}(\rho)=\inf_{\sigma\in\mcs}D(\rho,\sigma).
\eeq
It is immediate to check that if $D$ is contractive, then $E_{D}$ is a weak entanglement measure, as it vanishes on separable states by definition, and it satisfies weak monotonicity:
\[
\begin{split}
E_D(\Lambda_\locc[\rho])&=\inf_{\sigma\in\mcs}D(\Lambda_\locc[\rho],\sigma)\\
		&\leq \inf_{\sigma\in\mcs}D(\Lambda_\locc[\rho],\Lambda_\locc[\sigma])\\
		&\leq \inf_{\sigma\in\mcs}D(\rho,\sigma)\\
		&=E_D(\rho),
\end{split}
\]
with the first inequality justified by the fact that the set of separable states is mapped into itself by any LOCC operation. The [MOA] condition can be satisfied if $D$ has some additional properties~\cite{PlenioVedral1998}. Moreover, one can argue that the infimum in \eqref{eq:distmeas} is actually a minimum that can be achieved by some -- but not necessarily unique -- separable state $\sigma^*$, i.e. $E_D(\rho)=D(\rho,\sigma^*)$.

One particular LOCC operation, that we will denote by $\Lambda^{\sigma}_p$, consists in the addition of noise in the form of a separable state $\sigma$ with some probability $p$, i.e.
\[
\Lambda^{\sigma}_p[\rho]=(1-p)\rho+p\sigma.
\]
We will need the following result.
\begin{lemma}
\label{lem:dist_ent}
Suppose $D$ is a convex and contractive distance. Given any state $\rho$, and any probability $p$, if $\sigma^*$ realizes the infimum for $\rho$ in \eqref{eq:distmeas}, then
\beq
E_D(\Lambda^{\sigma^*}_p[\rho])=(1-p)E_D(\rho).
\eeq
\end{lemma}
\begin{proof}
On the one hand,
\[
\begin{split}
E_D(\Lambda^{\sigma^*}_p[\rho])&=\inf_{\sigma\in\mcs}D(\Lambda^{\sigma^*}_p[\rho],\sigma)\\
			&\leq \inf_{\sigma\in\mcs}\Big((1-p)D(\rho,\sigma)+p D(\sigma^*,\sigma)\Big)\\
			&\leq (1-p)E_D(\rho)
\end{split}
\]
having used the convexity of $D$ and taking $\sigma=\sigma^*$.
On the other hand,
\[
\begin{split}
E_D(\Lambda^{\sigma^*}_p[\rho])&=\inf_{\sigma\in\mcs}D(\Lambda^{\sigma^*}_p[\rho],\sigma)\\
			&\geq \inf_{\sigma\in\mcs}\Big(D(\rho,\sigma)-D(\rho,\Lambda^{\sigma^*}_p[\rho])\Big)\\
			&\geq \inf_{\sigma\in\mcs}D(\rho,\sigma)-pD(\rho,\sigma^*)\\
			&=(1-p)E_D(\rho),
\end{split}
\]
thanks to the triangle inequality and the convexity of $D$.
\end{proof}

We are now in the position to derive lower and upper bounds for $\Ee^{(D)}$ in terms of the original measure $E$ and of the distance measure $E_D$.
 
\begin{teo}
\label{thm:bounds}
Let $E$ be a convex entanglement measure, and $D$ a convex contractive distance. Then $\Ee^{(D)}$ satisfies the relation
\beq
\label{eq:bounds}
\min_{\substack{\tau\,\textrm{s.t.}\\
E_D(\tau)=E_D(\rho)-\epsilon}}
E(\tau)\leq\Ee^{(D)}(\rho)\leq \Big(1-\frac{\epsilon}{E_D(\rho)}\Big)E(\rho)
\eeq
\end{teo}
\begin{proof} 
We will first derive the upper bound, then the lower one. Both bounds rely on the geometric intuition represented in Fig.~\ref{fig:bounds}.
\begin{figure}[h]
 \includegraphics[width=0.45\textwidth]{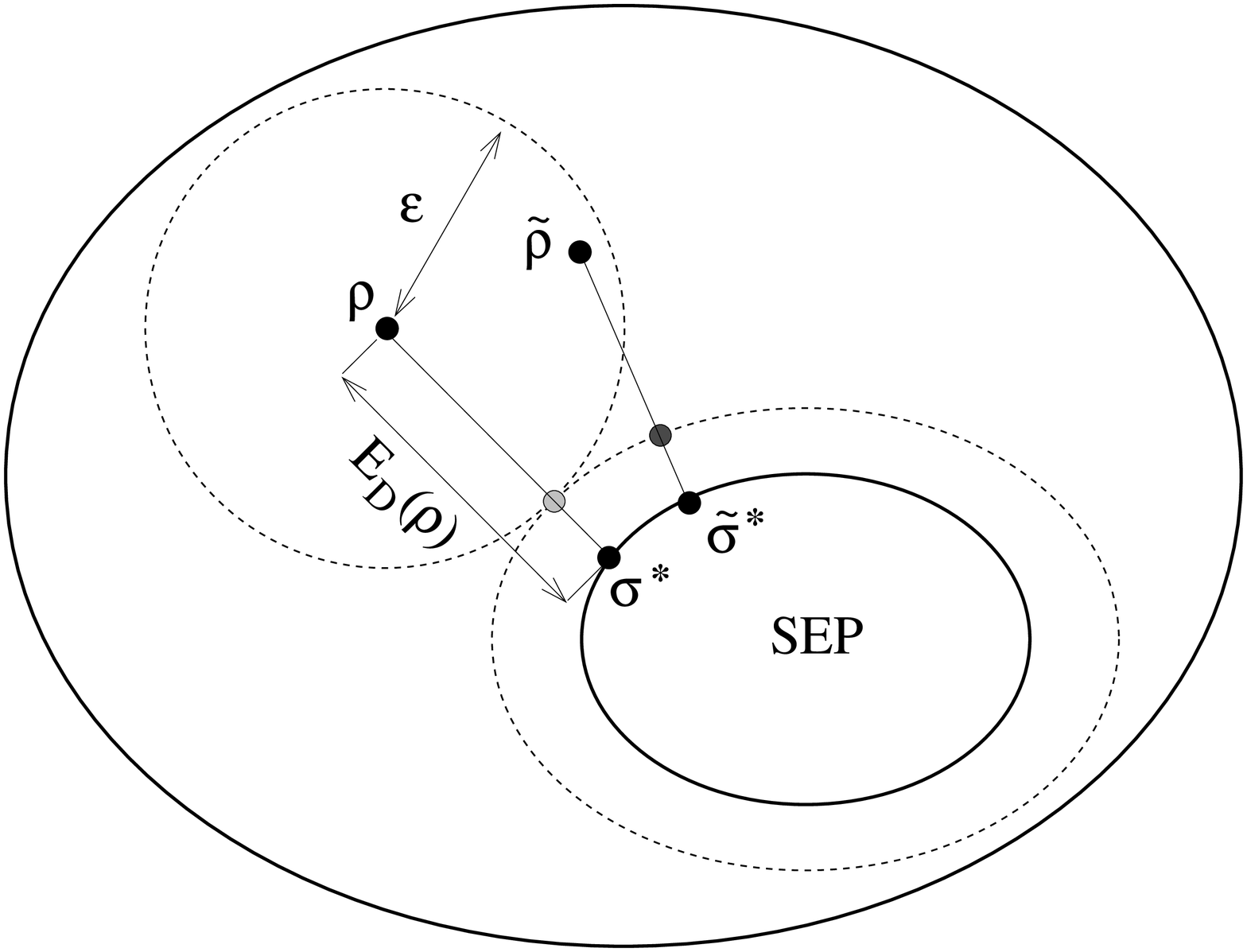}
\caption{Proof of the bounds \eqref{eq:bounds}. The largest ellipse denotes the whole set of states, while the smallest describes the set of separable ones. The dashed circle represents the ball of radius $\epsilon$ around $\rho$, while the dashed ellipse corresponds to the surface of states $\tau$ satisfying $E_D(\tau)=E_D(\rho)-\epsilon$, i.e. the states that are exactly ($E_D(\rho)-\epsilon$) far away --as measured by the distance $D$-- from separable states. The dark-gray circle and the light-gray one correspond to the states which help to prove the lower and upper bounds, respectively. See the main text for more details.}
\label{fig:bounds}
\end{figure}

By definition, $\Ee(\rho)\leq E(\tau)$ for all $\tau\in B_\epsilon(\rho)$. In particular, $\Ee(\rho)\leq E(\Lambda_{\locc}[\rho])$, with $D(\Lambda_{\locc}[\rho],\rho)\leq \epsilon$. Let us consider $\Lambda_{\locc}=\Lambda^{\sigma}_p$, for some probability $p$ and separable state $\sigma$.
Since $E$ and $D$ are assumed to be convex, we have $E(\Lambda^{\sigma}_p[\rho])\leq (1-p)E(\rho)$ and $D(\Lambda^{\sigma}_p[\rho],\rho)\leq p D(\sigma,\rho)$. Thus, $D(\Lambda^\sigma_p[\rho],\rho)\leq \epsilon$ for $p\leq \epsilon/D(\sigma,\rho)$.
It follows that:
\[
\begin{split}
\Ee(\rho)&\leq
\min_{\substack{\Lambda^\sigma_p\, \textrm{s.t.}\\
            		D(\Lambda^\sigma_p[\rho],\rho)\leq \epsilon}}
E(\Lambda^\sigma_p[\rho])\\
&\leq \min_{\substack{\Lambda^\sigma_p\, \textrm{s.t.}\\
            		D(\Lambda^\sigma_p[\rho],\rho)\leq \epsilon}}
(1-p)E(\rho)\\
&\leq \min_{\sigma}\Big(1-\frac{\epsilon}{D(\rho,\sigma)}\Big)E(\rho)\\
&\leq \Big(1-\frac{\epsilon}{E_D(\rho)}\Big)E(\rho),
\end{split}
\]
where $E_D(\rho)=\min_{\sigma\in\mcs}D(\rho,\sigma)$ is the entanglement measure associated to the distance $D$, and the third inequality is due to a restriction of the minimum to the case where we fix the value $p=\epsilon/D(\sigma,\rho)$.

On the other hand, we have $\Ee(\rho)=E(\tilde{\rho})\geq E(\Lambda_{\locc}[\tilde{\rho}])$ for any $\tilde{\rho}\in B_\epsilon(\rho)$ that realizes the minimum in the definition of $\Ee$. Let us consider a separable state $\tilde{\sigma}$ optimal for $E_D(\tilde{\rho})$. Because of the triangle inequality, we have
\beq
\label{eq:s_pos}
\begin{split}
E_D(\tilde{\rho})&=D(\tilde{\rho},\tilde{\sigma})\\
								 &\geq D(\rho,\tilde{\sigma})- D(\tilde{\rho},\rho)\\
							   &\geq E_D(\rho) - \epsilon.
\end{split}
\eeq
as $\tilde{\sigma}$ may not be optimal for $E_D(\rho)$. Let us take
\[
s=1-\frac{E_D(\rho) - \epsilon}{E_D(\tilde{\rho})},
\]
which satisfies $0\leq s \leq 1$ because of \eqref{eq:s_pos}. Then $E_D(\Lambda^{\tilde{\sigma}}_s[\tilde{\rho}])=E_D(\rho) - \epsilon$  thanks to Lemma \ref{lem:dist_ent}.
Therefore,
\[
\begin{split}
\Ee(\rho)&=E(\tilde{\rho})\\
				&\geq E(\Lambda^{\tilde{\sigma}}_s[\tilde{\rho}])\\
				&\geq \min_{\substack{\tau\, \textrm{s.t.}\\
						E_D(\tau)=E_D(\rho)-\epsilon}}
						E(\tau).
\end{split}
\]
as $\Lambda^{\tilde{\sigma}}_s$ is an LOCC operation and $E$ is an LOCC monotone.
\end{proof}


Let us consider the distance-based entanglement measure $E_D$, and let us take its $\epsilon$-generalization to be based on the same distance $D$ entering in its definition \eqref{eq:distmeas}. Then, Theorem \ref{thm:bounds} gives immediately that $(E_D)^{(D)}_\epsilon(\rho)=E_D(\rho)-\epsilon$.

\subsection{Relative entropy}

We have defined the $\epsilon$-version $\Ee^{(D)}(\rho)$ of a given entanglement measure $E(\rho)$ in terms of a distance $D$. The latter essentially provides the means to define a ball around the state $\rho$. It is possible to consider $\epsilon$-versions based on general functions, not necessarily distances, which allow us to define such a kind of ball. In particular, we may think of using relative entropy upon which one can define a measure of entanglement as in \eqref{eq:distmeas}, obtaining the so-called relative entropy of entanglement \cite{PlenioVedral1998}, which has a wide range of applications in quantum information theory~\cite{reviewent,reviewplenio}. The relative entropy of $\rho$ with respect to $\sigma$ is defined as
\[
S(\rho||\sigma)=\Tr \rho \log \rho - \Tr \rho \log \sigma,
\]
where one conventionally assumes that $S(\rho||\sigma)=+\infty$ if the support of $\rho$ is not included in the support of $\sigma$. Relative entropy is not a distance, since it is not symmetric and it does not satisfy the triangle inequality, nevertheless it shares three key features with the distance measures we have considered so far: it is contractive under CPT maps, it is jointly convex, and $S(\rho||\sigma)=0$ if and only if $\rho=\sigma$ \cite{NC}. It is thus clear that most of the proofs given in the case of $\Ee^{(D)}$ -- with the exception of the second parts of Theorem \ref{thm:continuity}, namely the continuity in $\rho$, and of Theorem \ref{thm:bounds}, i.e. the lower bound on  $\Ee^{(D)}$, which use the triangle inequality -- hold also for
\beq
E^{(R)}_\epsilon(\rho)=\min\{ E(\sigma)~|~\sigma\in B^{(R)}_\epsilon(\rho)\},
\eeq
with $B^{(R)}_\epsilon(\rho)=\{\sigma~|~S(\sigma||\rho)\leq\epsilon\}$. In particular, this implies that the upper bound
\[
(E_R)_\epsilon^{(R)}(\rho)\leq E_R(\rho)-\epsilon
\]
is still valid.

\section{Conclusions}
\label{sec:conclusions}

In this paper we have introduced the concept of $\epsilon$-measure of entanglement, which can be associated to any pre-existent measure. Such a quantity aims at quantifying the entanglement contained in a state of which we have only partial knowledge such as, for example, in the case of an imperfect preparation. The $\epsilon$-measure of a quantum state can thus be interpreted as the minimum ``guaranteed'' entanglement contained in the actually prepared state, given that we only know that it is $\epsilon$-close to the ideal target state.

On the one hand, we have show that the $\epsilon$-version of any entanglement measure is still an entanglement measure satisfying weak monotonicity. On the other hand, no $\epsilon$-measure satisfies monotonicity on average under LOCC operations. These two facts could lead to a better understanding of the physical meaning of the difference between these two kinds of monotonicity. Furthermore, we proved that the $\epsilon$-version of a convex entanglement measures is continuous and thus can also be seen as a ``smoothing'' of the original quantity, which could be non-continuous.

We believe that the newly introduced quantities will play a significant role in any context where it is necessary to take uncertainty in the preparation or in the knowledge of a state into account. This has already been the case for the study of properties of universal resources for measurement-based quantum computing~\cite{univprl,univlong} in the approximate scenario~\cite{univinprep}.  

We thank O. G{\"u}hne, B. Kraus, and M. Horodecki for discussions. We acknowledge support by the Austrian Science Fund (FWF), in particular through the Lise Meitner Program, and the EU (OLAQUI,SCALA,QICS).

\end{document}